\documentclass[aps,showpacs,superscriptadress,amsmath,preprint]{revtex4}
\usepackage{dcolumn,graphics,epsf}
\usepackage{color}

\newcommand{\bfg}[1]{\mbox{\boldmath $#1$}}

\begin{document}

\title{Direct above-threshold ionization in intense few-cycle laser pulses:
structures in the angle-integrated energy spectra}

\author {C.\ C.\ Chiril\u{a}}
\affiliation{Institute of High Performance Computing, Agency for Science,
Technology and Research, 1 Fusionopolis Way, \#16-16 Connexis, Singapore 138632}
\author {R.\ M.\ Potvliege}
\affiliation{Joint Quantum Centre (JQC) Durham-Newcastle, Department of Physics, Durham University, Durham
  DH1 3LE, United Kingdom}

\begin{abstract}
This paper concerns the theory of non-recollisional
ionization or detachment
of atoms or ions by intense few-cycle pulses. It is shown that 
in certain conditions of pulse duration,
peak intensity and carrier-envelope phase,
the ionization probability integrated over ejection angle varies
almost periodically with energy, with a period roughly equal
to the photon energy for slow enough outgoing electrons.
This modulation is found both in calculations based on
the strong field approximation (SFA) and in
{\it ab initio} time-dependent calculations. It is explained as
resulting from
the interference between the contributions of different saddle times
of the modified classical action.
Methods for efficiently calculating the SFA ionization amplitude 
beyond the usual saddle point approximation
are also discussed.
\end{abstract}

\pacs{32.80.Rm,42.50.Hz} 
\maketitle

\section{Introduction}

Ever since the discovery of above-threshold ionization (ATI)  \cite{Agostini},
the energy spectrum of the electrons ejected from atoms or ions 
exposed to an intense laser pulse has proved
rich in interesting features \cite{Beckerrev}.
These include the suppression of the lowest ATI peaks in long pulses,
which arises from the
ponderomotive acceleration of the outgoing electron \cite{Petite},
prominent Stark-shift induced resonances 
in short pulses
\cite{Freeman},
the recollision plateau, which extends the spectrum well beyond the classical
cutoff for direct ionization \cite{Paulus},
and the low energy and very low energy structures
recently found in ionization by
 ultrashort infrared pulses \cite{lowen}.

The theory of multiphoton ionization has progressed in parallel with these
discoveries,
through a combination
of {\it ab initio} time-dependent calculations and of
analyses based on the strong
field approximation (SFA) or on the Floquet theory and other approaches
\cite{Book}. Of particular note in the context of the present
work is Keldysh's theory \cite{Keldysh}, a length-gauge
formulation of the SFA which,
since its inception fifty years ago, has been at the basis of a large
fraction of the theoretical work on strong field physics.
This formulation predicts accurate ionization probabilities 
for detachment from negative ions \cite{Kuchiev}. It is also
both qualitatively and quantitatively correct for ionization from atoms,
provided the Coulomb interaction between the active electron
and the residual ion is properly taken into account \cite{CCRMP}.

In this article
we study another feature of ionization in intense ultra-short
laser
pulses, namely an almost periodic modulation marking the angle-integrated
energy spectrum in appropriate conditions of pulse duration, peak intensity
and carrier-envelope phase. Although this modulation is readily found both
in SFA calculations and in {\it ab initio} calculations, we are not aware
that it has been discussed previously \cite{thesis}. However,
like various other strong field phenomena,
it can be traced, through the SFA, to the interplay between different
saddle times of the modified classical action.
We concentrate on the low energy end of the ionization spectrum, where
this modulation is clearest. As is well known, this part of the spectrum
is dominated by direction ionization. Recollision of the detached electron
with the residual ion plays no role here, and is therefore neglected
in our analysis.

The theoretical background to the work is outlined in Section \ref{sect:theory}.
The results are presented and discussed in Section \ref{sect:results}.
Technical issues concerning the calculation of the ionization amplitude
within the SFA are briefly considered in the Appendix --- namely avoiding
the spurious 
contributions made to the ionization amplitude
by the end points of the integral
defining it,
computational methods bypassing saddle point
integration, and improving the accuracy of the usual (second order) saddle
point method.
Atomic units are assumed throughout this article,
except where specified otherwise.

\section{Theory}
\label{sect:theory}

We work within the dipole approximation and
describe the laser pulse by a spatially homogeneous vector
potential ${\bf A}(t)$ and a spatially homogeneous electric field ${\bf F}(t) =
-\partial_t {\bf A}(t)$. 
Specifically, we set
\begin{equation}
{\bf A}(t) = (F_0/\omega) \chi(t)  \hat{\bfg{\epsilon}} \sin
(\omega t +\varphi),
\label{eq:pulse}
\end{equation}
where $\hat{\bfg{\epsilon}}$ is a unit vector (we assume linear
polarization), $\chi(t)$ is a function
defining the pulse's intensity profile,
and $\varphi$ is an arbitrary phase \cite{cep}.
We assume that $\chi(t)$ peaks at $t=0$.
Most of the results presented below are
calculated 
for pulses with a half-period $\cos^2$ amplitude envelope 
encompassing an integer number 
of optical cycles, for which
\begin{equation}
\chi(t)=\begin{cases}\cos^2 \left(\displaystyle{\omega t \over 2 n_{\rm c}}\right) &
 - n_{\rm c}\pi/\omega \leq t \leq n_{\rm c}\pi/\omega \cr
0 & t <  - n_{\rm c}\pi/\omega \quad \mbox{or} \quad t >  n_{\rm c}\pi/\omega,
\end{cases}
\end{equation}
where $n_{\rm c}$ is the number of optical cycles encompassed by the
pulse. Such pulses have the desirable
property of not imparting an unphysical displacement or drift momentum to
a free electron \cite{Dondera}.
We also consider pulses with a sech amplitude profile (sech$^2$ in
intensity). In either case,
both $|{\bf A}(t)|$ and $|{\bf F}(t)|$ are negligibly small, if not exactly zero,
before a certain time
$t_{\rm i}$ and after a certain time $t_{\rm f}$.

For simplicity, we assume that the atom has
only one active electron and is initially
in a certain bound state
with wave function $\Phi_0({\bf r},t) = \phi_0({\bf r})\exp(I_p t)$.
In Keldysh's formulation of the strong field approximation, the probability
amplitude for the photoelectron to have a momentum 
${\bf p}$ at times $t \geq t_{\rm f}$ is then
\begin{equation}
A^{({\rm K})}_{{\bf p}0} =
-i\int_{t_{\rm i}}^{t_{\rm f}}{d}t\int {d}{\bf r}\,
\Psi_{\bf p}^*({\bf r},t)
\left[{\bf r}\cdot{\bf F}(t)\right]\Phi_0({\bf r},t),
\label{eq:Adef}
\end{equation}
within an irrelevant phase factor. The wave function
$\Psi_{\bf p}({\bf r},t)$ is the Volkov wave 
\begin{equation}
\Psi_{\bf p}({\bf r},t)= {1 \over (2\pi)^{3/2}}
\exp\left[ i\bfg{\pi}({\bf p},t)\cdot{\bf r} -
\frac{i}{2}\int_{t_{\rm i}}^t\!
dt' \bfg{\pi}^2({\bf p},t)\right],
\label{eqn:Volkov}
\end{equation}
where $\bfg{\pi}({\bf p},t)$ denotes
the kinematical momentum of the electron:
\begin{equation}
\bfg{\pi}({\bf p},t) = {\bf p} + {\bf A}(t).
\label{eq:kine}
\end{equation}
In this formulation,
the interaction between the photoelectron and the ionic core
is treated exactly in the initial state of the system but is
otherwise completely neglected.
As is well known, the effect of
this long range interaction on the motion of the electron
during the tunnelling stage of the ionization process
can be taken into account semiclassically, and doing so
brings the predictions of the theory into much closer
agreement with experiment. For a stationary
laser field of electric field amplitude $F_0$, the correction amounts to
multiplying $\Psi_{\bf p}({\bf r},t)$
by the factor \cite{Krainov}
\begin{equation}
I(r)=\left({4 I_p \over F_0}{1\over  r}\right)^{Z/\kappa},
\end{equation}
where $Z$ is the charge of the residual ion and $\kappa=(2I_p)^{1/2}$.
Although derived for
a stationary field, this correction, with $F_0$ taken to the
peak electric field amplitude, has been shown to be effective for 
ultra short laser pulses \cite{CCRMP}.
Rather than the Keldysh
amplitude (\ref{eq:Adef}),
we thus work with the ``tunnelling corrected" amplitude
\begin{equation}
A_{{\bf p}0} =
-i\int_{t_{\rm i}}^{t_{\rm f}}{d}t\int {d}{\bf r}\,
\Psi_{\bf p}^*({\bf r},t)
I(r) \left[{\bf r}\cdot{\bf F}(t)\right]\Phi_0({\bf r},t).
\label{eqn:Adefcorr}
\end{equation}
Given the normalization of the Volkov wave (\ref{eqn:Volkov}),
the density of
probability that an electron is detached by the pulse
with a final kinetic energy $E=p^2/2$ is
\begin{equation}
P(E) = 2\pi \int_0^\pi\, 
P(E,\theta) \, \sin\theta\,{d}\theta,
\end{equation}
with $\theta$ the angle between the momentum ${\bf p}$ and the polarization
vector $\hat{\bfg{\epsilon}}$ and
\begin{equation}
P(E,\theta) = p\, \left|A_{{\bf p}0}\right|^2.
\end{equation}

Eq.\ (\ref{eqn:Adefcorr}) can also be written in the form
\begin{equation}
A_{{\bf p}0} =
-{1 \over (2\pi)^{3/2}}\int_{t_{\rm i}}^{t_{\rm f}}{d}t\int {d}{\bf r}\,
\left({\partial\; \over \partial t} \exp[-i\bfg{\pi}({\bf p},t)\cdot{\bf r}]
\right)
I(r)\, \phi_0({\bf r}) \exp[iS({\bf p},t)],
\label{eqn:Amod}
\end{equation}
with
\begin{equation}
S({\bf p},t) = 
\frac{1}{2}\int_{t_{\rm i}}^t\!
\bfg{\pi}({\bf p},t')^2 dt' + I_pt.
\label{eq:Sdef}
\end{equation}
Upon integrating by parts, we thus have
\begin{equation}
A_{{\bf p}0} =
-{1 \over (2\pi)^{3/2}}
\exp[iS({\bf p},t)]M_{{\bf p}0}(t)\Big\vert_{t_{\rm i}}^{t_{\rm f}}
+{i \over (2\pi)^{3/2}}\int_{t_{\rm i}}^{t_{\rm f}}
\exp[iS({\bf p},t)] \,  S^\prime({\bf p},t)
M_{{\bf p}0}(t) dt
\label{eqn:Amod2}
\end{equation}
where $S^\prime({\bf p},t)$ is the derivative of
$S({\bf p},t)$ with respect to time and
\begin{equation}
M_{{\bf p}0}(t) = \int
\exp[-i\bfg{\pi}({\bf p},t)\cdot{\bf r}]
I(r)\, \phi_0({\bf r}) d{\bf r}.
\end{equation}
As we will soon see, the boundary
terms appearing in Eq.\ (\ref{eqn:Amod2}) are
exactly cancelled by opposite contributions
from the end-points of the integral.
We note from Eq.\ (\ref{eqn:Adefcorr})
that in fact the ionization amplitude $A_{{\bf p}0}$ does not depend
on the precise values of the initial and final times
$t_{\rm i}$ and $t_{\rm f}$, as long as the pulse's electric field 
is effectively zero at and around $t_{\rm i}$ and at $t_{\rm f}$
(as should be expected on physical grounds --- the ionization probability
cannot depend on how the field varies at times where it is too weak to
affect the atom).
Mathematically, there is no dependence on $t_{\rm i}$ and $t_{\rm f}$
only if ${\bf E}(t)$ and all the derivatives of ${\bf E}(t)$
vanish at these two times \cite{boundary2}.
This condition is not met by the model
of pulses commonly used in calculations, which might have practical
consequences if pulses with an excessively fast turn on and turn off are 
considered (this issue is considered further in the Appendix but is normally not
problematic in applications to realistic cases).

More specifically, we represent the initial state of the atom by an s-orbital
and, following \cite{PPT}, set
\begin{equation}
\phi_0({\bf r}) \equiv 2 \kappa^{3/2} C_{\kappa 0} (\kappa r)^{(Z/\kappa)-1}
\exp(-\kappa r)/\sqrt{4\pi},
\end{equation}
where $C_{\kappa 0}$ is the asymptotic coefficient for the
species considered in the definition of \cite{popovreview}.
Accordingly, the product
$S^\prime({\bf p},t) M_{{\bf p}0}(t)$ reduces to
$(4\pi \kappa)^{1/2}(4I_p\kappa/F_0)^{Z/\kappa}C_{\kappa 0}$, and
\cite{velgauge}
\begin{equation}
A_{{\bf p}0} = i
{(2\kappa)^{1/2} C_{\kappa 0}  \over 2 \pi}
\left({4I_p \kappa \over F_0}\right)^{Z/\kappa}
\left(
\int_{t_{\rm i}}^{t_{\rm f}} \exp[iS({\bf p},t)]dt
 - \left. {\exp[iS({\bf p},t)] \over iS^\prime({\bf p},t)}
\right|_{t_{\rm i}}^{t_{\rm f}}\right).
\label{eqn:Amod3}
\end{equation}
In this article we present results for detachment from the 
ground state of an He$^+$ ion or an hydrogen atom,
for which $C_{\kappa 0} = 1$, and for ionization from the ground state of 
neutral helium, 
for which it is appropriate to take $C_{\kappa 0}=0.993$
\cite{popovreview}.

The integral appearing in the right-hand side of Eq.\ (\ref{eqn:Amod3})
is amenable to saddle point integration, which is the usual way of
calculating the ionization amplitude in the strong field approximation.
Within this approach,
\begin{equation}
\int_{t_{\rm i}}^{t_{\rm f}}
\exp[iS({\bf p},t)] \, 
 dt \approx 
\sum_{j} \sqrt{2\pi i \over S^{\prime\prime}({\bf p},t_{j})}
\, \exp[iS({\bf p},t_{j})]
 + \left. {\exp[iS({\bf p},t)] \over iS^\prime({\bf p},t)}
\right|_{t_{\rm i}}^{t_{\rm f}}+\ldots
\label{eqn:s}
\end{equation}
where the times $t_j$ are the complex values of $t$ at which
$S'({\bf p},t)=0$.
The first term in the right-hand side of Eq.\ (\ref{eqn:s}) is
the contribution to the integral
of the saddle times $t_j$, while the second term
is the dominant contribution of the end-points $t_{\rm i}$ and
$t_{\rm f}$ of the integration contour 
(dominant in the sense of an asymptotic analysis,
see, e.g., \cite{Bleistein}). Depending on the pulse,
the second term can be large, even much larger than the first term;
however, as shown by Eq.\ (\ref{eqn:Amod3}),
it is exactly cancelled by
the boundary terms arising from the integration by parts. The remainder,
not written down explicitely in the equation, is the sum of the
higher-order contributions
of these two end-points and of the saddle times. Since the choice of $t_{\rm i}$ and $t_{\rm f}$
is arbitrary, it is appropriate to neglect the end-point contributions
altogether and write
\begin{equation}
A_{{\bf p}0} \approx
i{(2\kappa)^{1/2} C_{\kappa 0}  \over 2 \pi}
\left({4I_p \kappa \over F_0}\right)^{Z/\kappa}
\sum_{t_{j}} \sqrt{2\pi i \over S^{\prime\prime}({\bf p},t_{j})}
\, \exp[iS({\bf p},t_{j})].
\label{eqn:As}
\end{equation}

However, for maximum accuracy, we prefer not to use
the saddle point method to calculate
the energy spectrum.
Instead, we treat time as a complex variable
and numerically
integrate the function
$\exp[iS({\bf p},t)]$ over $t$
along a straight line path 
parallel to the real axis and passing through the saddle point with
the lowest positive imaginary part. This approach and other alternative methods 
for integrating this function over the duration of the pulse are discussed
in the Appendix.

\section{Results and Discussion}
\label{sect:results}

\begin{figure*}
\includegraphics{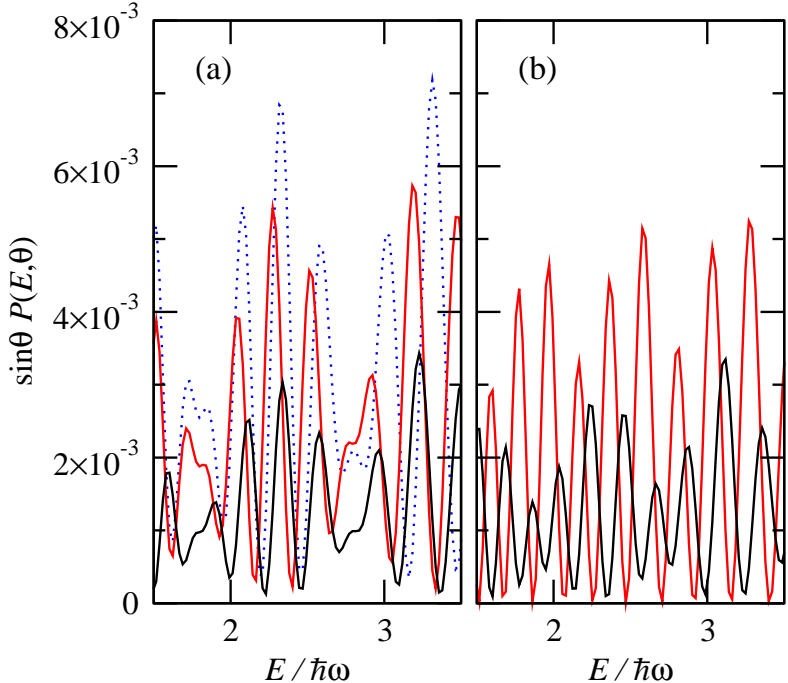}
\caption{(Color online) The probability of detachment from the ground
state of He$^+$ by a 4-cycle cos$^2$ pulse, (a) for $\varphi=0$, (b) for
$\varphi=\pi/2$. The carrier wavelength is 800 nm and
the peak intensity is about $5.6\times 10^{15}$ W cm$^{-2}$
($F_0=0.4$ a.u.\ exactly).
Solid black curves: $\theta=\pi/20$. Solid red curves: $\theta=\pi/10$.
Dotted blue curve (left panel only): $\theta=3\pi/20$.
}
\label{Fig:newfig1}
\end{figure*}
As is well known, the ionization probability predicted by
the SFA is generally an oscillating function
of the detachment energy $E$ and of the angle of emission $\theta$,
due to
interferences between the contributions of different saddle times $t_j$.
Examples of 
this oscillatory behavior are shown
in Fig.\ \ref{Fig:newfig1} for the case of a strong 800 nm 4-cycle
pulse interacting with an He$^+$ ion.
Panel (a)  illustrates the variation of $P(E,\theta)$
for a ``cosine-like" pulse ($\varphi=0$), panel (b)
for a ``sine-like" pulse ($\varphi=\pi/2$).
Comparing these two sets of results, it can be seen that the energies
at which $P(E,\theta)$ is maximal tend to vary less with the emission angle 
for $\varphi=0$ than for $\varphi=\pi/2$. In particular, for
$\varphi=0$ the peaks tend to come in groups concentrated
in the same ranges of energies
for all values of $\theta$.

\begin{figure*}
\includegraphics{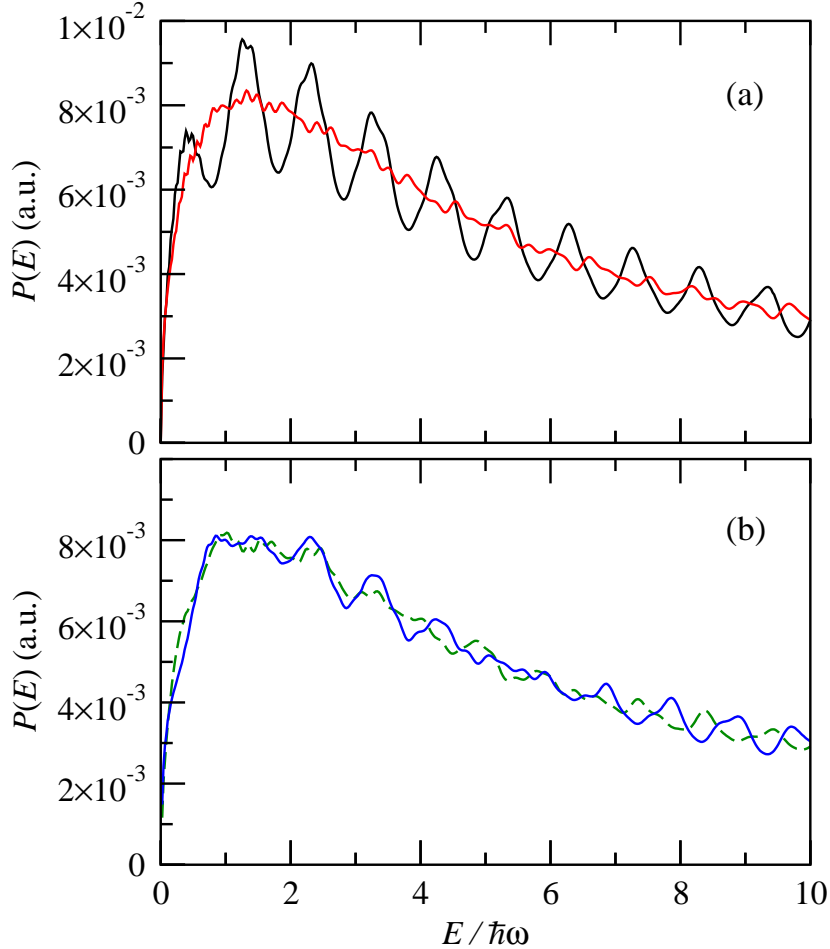}
\caption{(Color online) The angle-integrated probability of detachment from the ground
state of He$^+$ by a 4-cycle cos$^2$ pulse, (a) for $\varphi=0$ (black curve)
or $\varphi=\pi/2$ (red curve), (b) for
$\varphi=\pi/10$ (blue solid curve)
 or $\varphi=\pi/5$ (green dashed curve). As in Fig.\
\ref{Fig:newfig1}, the carrier wavelength is 800 nm and $F_0=0.4$ a.u.\
(about $5.6\times 10^{15}$ W cm$^{-2}$ peak intensity).
}
\label{Fig:newfig2}
\end{figure*}
This feature is more striking in the
angle-integrated spectra shown in Fig.\ \ref{Fig:newfig2}:
the propensity of 
$P(E,\theta)$ to be largest in the
same ranges of values of $E$ (almost) irrespective of $\theta$ results in 
broad, almost regularly spaced peaks modulating
the angle-integrated probability $P(E)$ when $\varphi \approx 0$.
For the pulse duration and intensity considered in Fig.\ \ref{Fig:newfig2}, these
peaks rapidly decrease in contrast
when $\varphi$ increases and they do not manifest 
for $\varphi=\pi/2$.
The peaks found for $\varphi \approx 0$, which are almost regularly spaced 
by the photon energy, 
are reminiscent of the well-known ATI
peaks observed in long-pulse experiments \cite{Book}. However, their
origin is different. Here ponderomotive scattering plays no role
and, as discussed below, these structures arise directly from the
way the modified classical
action $S({\bf p},t)$ varies with the angle of emission.
The spacing between the peaks found for few-cycle pulses is
actually energy-dependent, although this feature is not visible in
Fig.\ \ref{Fig:newfig2}

\begin{figure*}
\includegraphics{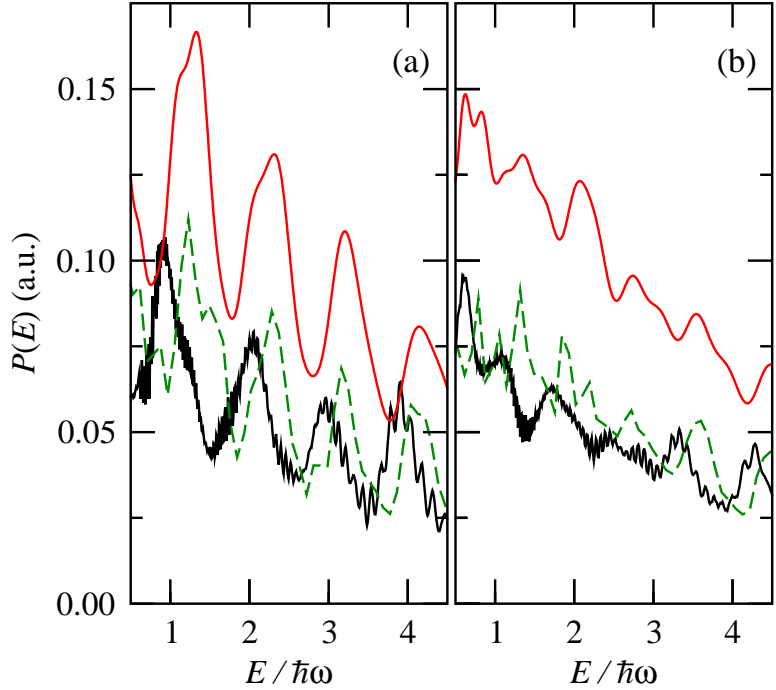}
\caption{(Color online) The angle-integrated probability of detachment from the ground
state of He$^+$ by a cos$^2$ pulse encompassing
exactly 4 optical cycles.
Here the carrier wavelength is 400 nm and the peak intensity
of the pulse is $1\times 10^{16}$ W cm$^{-2}$.
(a): $\varphi=0$.
(b): $\varphi=\pi/2$.
Solid black curves: Spectrum obtained by solving the time-dependent
Schr{\"o}dinger equation {\it ab initio}.
Solid red curves: Predictions of the strong field approximation.
Dashed green curves: The same as the solid black curves, but with the
spectrum calculated by projecting the wave function on plane waves.
}
\label{Fig:newfig6}
\end{figure*}
That these structures are not an artefact 
of the strong field approximation is shown by Fig.\ \ref{Fig:newfig6},
for a 400 nm pulse: the angle-integrated spectra obtained by
solving the time-dependent Schr\"odinger equation {\it ab initio} are very
similar to the SFA spectra and have the same periodic structure,
apart for an unimportant difference in overall
amplitude and a shift in the position of the peaks \cite{methodTD}.
A shift due to the Coulomb
interaction between the outgoing electron and the parent ion can be
expected --- see, e.g., \cite{CCRMP}. However, the predictions 
of the strong field approximation are well verified by the {\it ab initio}
calculation.
(Also shown in Fig.\ \ref{Fig:newfig6}, and represented by a dashed 
curve, is the spectrum
obtained by projecting the time-dependent 
wave function onto plane waves, at the end of the pulse, after this
time-dependent wave function has been orthogonalized to the initial state.
This approximate spectrum is a better comparison and is in better agreement
with the SFA spectrum
since it is not affected by the Coulomb force acting on the
outgoing electron after the end of the pulse.)

\begin{figure*}
\includegraphics{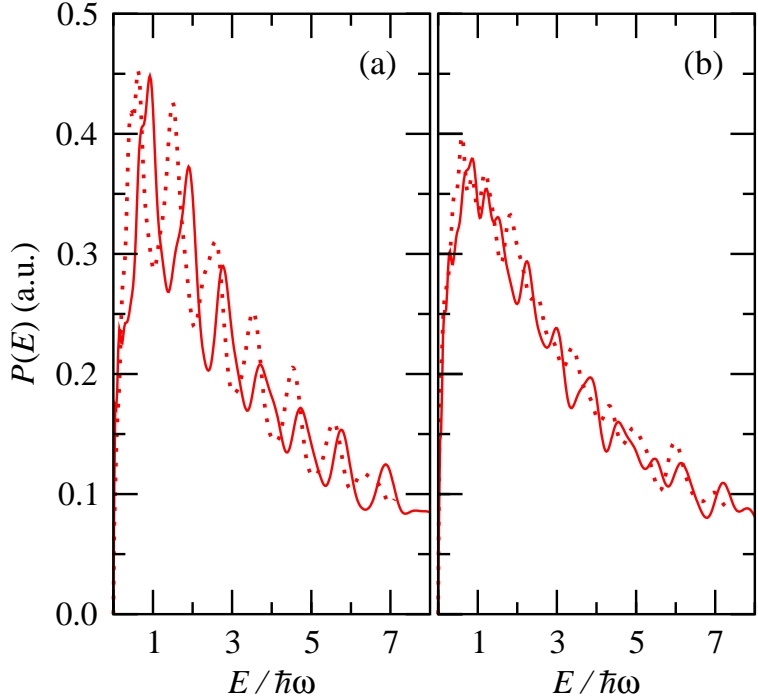}
\caption{(Color online) The probability of detachment from the ground state
of neutral helium by a few-cycle pulse. The carrier wavelength
is 800 nm. The pulse has a sech-profile in amplitude with a full
width at half maximum of 2 optical cycles. The peak intensity
is either $1.00\times 10^{15}$
W cm$^{-2}$ (solid curves) or $1.01\times 10^{15}$
W cm$^{-2}$ (dotted curves). (a): $\varphi=0$. (b): $\varphi=\pi/2$.
}
\label{Fig:newfig4}
\end{figure*}
However, observing these structures in a high-intensity experiment is likely
to be problematic, as the spectrum depends sensitively on the
parameters of the pulse. For example, the results of Fig.\ \ref{Fig:newfig4}
show that a mere 1\% change in the peak intensity,
from $1.00$ to $1.01\times 10^{15}$ W cm$^{-2}$, shifts the peaks very
significantly in the case of helium atoms ionized by a few-cycle 800 nm
pulse. (The amplitude envelope was taken to be a sech function in these
calculations, rather than a cos$^2$ function.)
Clearly, in experiments
using such strong fields, the peaks and troughs structure of the energy
spectrum would be averaged out by the
unavoidable spatial variation of the pulses'
intensity profile over the interaction region.

The origin of these peaks can be understood by analyzing how 
the ionization probability depends on the interference between the 
contribution of the different saddle times, in the approximation
where the ionization amplitude is given by Eq.\ (\ref{eqn:As}).
In this approximation,
\begin{equation}
P(E,\theta) \approx p\,C\,\left(
\sum_{j=1}^{n_{\rm s}}I_{jj}+2\sum_{j=1}^{n_{\rm s}-1}
\sum_{k=j+1}^{n_{\rm s}} I_{jk} \right),
 \label{eq:saddlesumcontrib1}
\end{equation}
where $n_{\rm s}$ is the number of saddle times making a non-negligible
contribution to $P(E,\theta)$, $C$ is 
a real positive constant,
and
\begin{equation}
I_{jk}=2\pi\,\mbox{Re}\left(
[S^{\prime\prime}({\bf p},t_{j})
 S^{\prime\prime*}({\bf p},t_{k})]^{-1/2}
\exp[iS({\bf p},t_{j})-iS^*({\bf p},t_{k})] \right).
\label{eq:Idefined}
\end{equation}
it is worth noting that $I_{jk}$ would be exactly proportional to
$\cos[
\mbox{Re}\,S({\bf p},t_j) -
\mbox{Re}\,S({\bf p},t_k)]$
if $S^{\prime\prime}({\bf p},t_{j})$ and $S^{\prime\prime*}({\bf p},t_{k})$
had no imaginary part.
The terms in $I_{jk}$ with $k\not= j$
may thus vary rapidly with $E$ and $\theta$ while
those with $k=j$ normally vary slowly.

In view of Eqs.\ (\ref{eq:pulse}), (\ref{eq:kine}) and (\ref{eq:Sdef}),
the saddle times $t_j$ and $t_k$ are solutions of the equation
\begin{equation}
(F_0/\omega)\chi(t)\sin(\omega t + \varphi) = -p_\parallel\pm i\sqrt{2I_p+p_{\perp}^2},
\label{eq:condition}
\end{equation}
where $p_\parallel = p \cos \theta$
and $p_\perp = p \sin\theta$.
(Only those solutions of this equation that have a positive imaginary part are
relevant in this context.)
For the small values of $E$ we are considering here,
the complex values of $t$ obtained by solving Eq.\ (\ref{eq:condition})
are sufficiently close to the real values of $t$ at which 
${\bf A}(t)=0$ that each relevant solutions can be sought by
expanding ${\bf A}(t)$ in powers of the difference $(t-t_0)$,
where $t_0$ is the zero of ${\bf A}(t)$ closest to the saddle time
considered. Doing so and limiting oneself to terms of second order
in $(t-t_0)$ yields
\begin{equation}\label{eq:interf}
\begin{split}
\mbox{Re}\,S({\bf p},t_j) -
\mbox{Re}\,& S({\bf p},t_k)
\approx \\
&(E+I_p)(t_{0j}-t_{0k}) + 
{1 \over 2}\int_{t_{0j}}^{t_{0k}} A^2(t) dt
+ a_{jk}p_\parallel 
\\ &  +p_\parallel\left(I_p+
\frac{p^2+2p_\perp^2}{6}\right)
\left[\frac{1}{E(t_{0k})}-\frac{1}{E(t_{0j})}\right],
\end{split}
\end{equation}
where
$t_{0j}$ and $t_{0k}$ are the real solutions of the equation
${\bf A}(t)=0$
closest to the complex saddle times $t_j$ and $t_k$ and
\begin{equation}
a_{jk}=\int_{t_{0j}}^{t_{0k}} 
 \hat{\bfg{\epsilon}}\cdot{\bf A}(t) dt.
\end{equation}
We stress that Eq.\ (\ref{eq:interf}) 
applies only for low
momenta of the ejected electron, which is the part of the spectrum
we focus on in this work. 

The difference $\mbox{Re}\,S({\bf p},t_j) -
\mbox{Re}\,S({\bf p},t_k)$
thus
depends on the angle of ejection $\theta$ primarily
through a term proportional to the integral $a_{jk}$ and a term
proportional to the difference $1/E(t_{0k})-1/E(t_{0j})$.
Hence, the contribution to $P(E,\theta)$ of those
pairs of saddle times for which
$a_{jk} \approx 0$ together with
$E(t_{0j}) \approx E(t_{0k})$ varies little with the ejection angle
$\theta$.

In the definition of the pulse adopted in this work, where the pulse envelope
is symmetric and peaks at $t=0$, such pairs of saddles exist for $\varphi=0$:
for this carrier-envelope phase, ${\bf A}(t)=0$ at $t_{01}=-\pi/\omega$,
$t_{02}=0$ and 
$t_{03}=\pi/\omega$, besides other values of $t$ of lesser relevance 
for a few-cycle pulse
(because the corresponding electric field
is somewhat weaker than at $t_{01}$, $t_{02}$ and $t_{03}$).
Let us call $t_1$, $t_2$ and $t_3$ the complex saddle times
closest to, respectively,
$t_{01}$, $t_{02}$ and $t_{03}$.
The contribution of $t_1$ and $t_3$ to $P(E,\theta)$
is (almost) angle-independent
since
$a_{13}=0$ and $E(t_{01})=E(t_{03})$. 
The interference between these two saddle times is constructive 
rather than destructive,
giving a peak in the spectrum, at the values of $E$ for which
$E \approx E_N$, where
\begin{equation}
E_N=N\omega-\left[I_p+{\omega \over 4\pi}
\int_{-\pi/\omega}^{\pi/\omega}A^2(t) dt\right]
\label{eqn:Eq23}
\end{equation}
with $N$ an integer \cite{energyconservation}.
The angular distribution will also depend on interferences between the 
contributions of $t_2$ and either $t_1$ or $t_3$.
However, since the corresponding values of $a_{jk}$ and
$1/E(t_{0k})-1/E(t_{0j})$ are non-zero, these contributions oscillate
rapidly with $\theta$ and hardly manifest in the
angle-integrated spectrum.

Turning to the case of $\varphi\approx \pi/2$, the most
relevant saddle times for few-cycle pulses are 
$t_1 \approx t_{01}\equiv -3\pi/2\omega$,
$t_2 \approx t_{02}\equiv -\pi/2\omega$,
$t_3 \approx t_{03}\equiv \pi/2\omega$ and
$t_4 \approx t_{04}\equiv 3\pi/2\omega$. 
For the peak intensity and pulse duration of Fig.\ \ref{Fig:newfig2},
however, the electric field is too weak at $t_1$ and $t_4$ for these 
saddles times to play an important role, and only $t_2$ and $t_3$ need to be
considered. Since $a_{23}\not=0$ and $E(t_{02})\not= E(t_{03})$, $P(E,\theta)$
oscillates rapidly both as a function of $E$ and of $\theta$,
and the resulting angle-integrated spectrum is almost structureless
[Fig.\ \ref{Fig:newfig2}(a)].

\begin{figure*}
\includegraphics{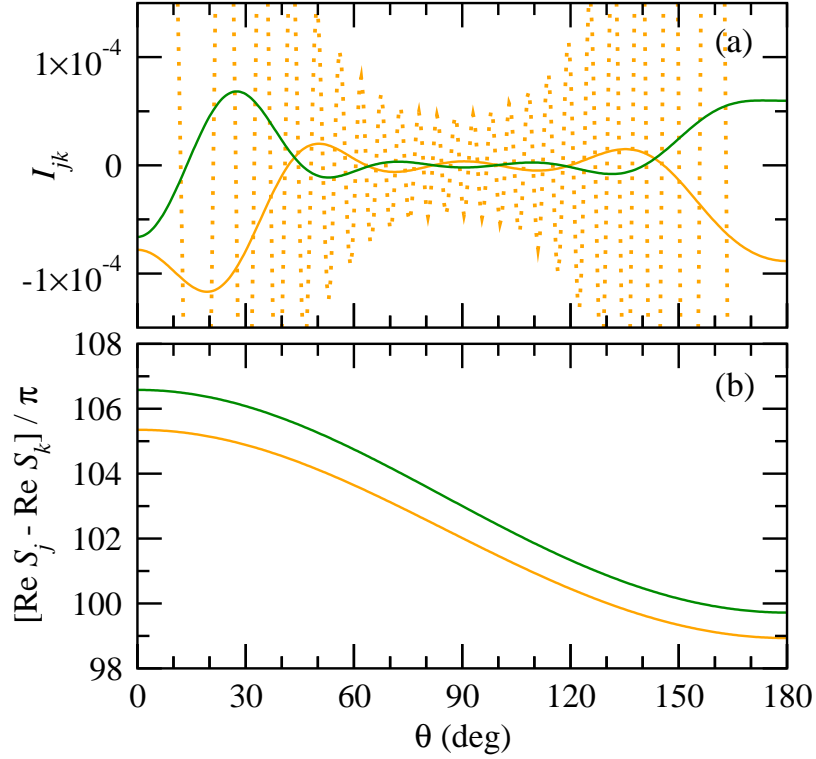}
\caption{(Color online)
Contribution of individual saddle times to the ionization amplitude
for the same system as in Fig.\ \ref{Fig:newfig4}(b), 
either for $E = 3.4\,\hbar\omega$ (light orange curves)
or $E = 3.9\, \hbar\omega$ (dark green curves) (the pulse peak intensity
is $1\times 10^{15}$ W cm$^{-2}$). (a): Difference between
the real part of the modified classical action at two different saddle times, divided
by $\pi$. (b): The quantity $I_{jk}$ defined by Eq.\ (\ref{eq:Idefined}).
Solid curves: 
$t_j \approx -3\pi/2\omega$ and $t_k \approx \pi/2\omega$.
Dotted curve:
$t_j \approx -\pi/2\omega$ and $t_k \approx \pi/2\omega$.
}
\label{Fig:newfig5}
\end{figure*}
Other saddle times can become significant in longer pulses 
or closer to saturation. For example, the
peaks and troughs visible in Figs.\ \ref{Fig:newfig6}
and \ref{Fig:newfig4} in the 
angle-integrated angular distribution for $\varphi=\pi/2$ arise from 
contributions from the saddle times $t_1$ and $t_4$ defined in the
previous paragraph, besides $t_2$ and $t_3$. 
The structures found in the case
of Fig.\ \ref{Fig:newfig4}(b) are analyzed in
Fig.\ \ref{Fig:newfig5}. Part (a) of
the latter shows how $I_{13}$ and $I_{23}$
vary with the ejection angle $\theta$ at either $E=3.4\, \hbar\omega$
(where the angle-integrated spectrum has a peak) or $3.9\, \hbar\omega$
(the adjacent trough). 
$I_{13}$, represented by the solid curves, oscillates much
less than $I_{23}$ (the dotted curve) both
because $|a_{13}| \ll |a_{23}|$ and
because $E(t_1) \approx E(t_3)$ whereas 
$E(t_2) = - E(t_3)$ ($a_{13}$ and $E(t_1) - E(t_3)$ would be zero if
the field had a constant intensity). As seen from the figure,
$I_{13}$ keeps the same sign
in the angular regions where this term contributes most to the ionization
probability \cite{explanation}. Positive values give a
peak in the angle integrated spectrum, and negative values a trough. 
Increasing $E$ leads to
a near-periodic succession of peaks and troughs
because, as seen from Fig.\ \ref{Fig:newfig5}(b), 
$\mbox{Re}\,[S({\bf p},t_1) -
S({\bf p},t_3)]$ increases almost uniformly with $E$. $I_{13}$ oscillates
between positive and negative values as $\mbox{Re}\,[S({\bf p},t_1) -
S({\bf p},t_3)]$ sweeps through half integer multiples of $\pi$
 (or thereabout).

\begin{figure*}
\includegraphics{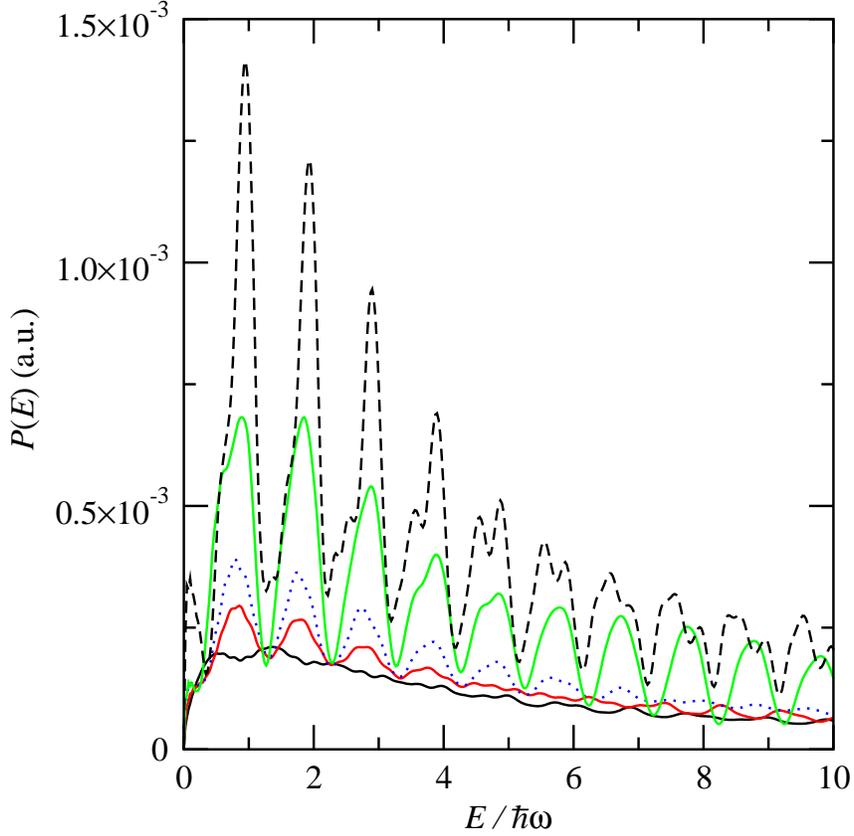}
\caption{(Color online) The angle-integrated probability of detachment from the ground
state of He$^+$ by a cos$^2$ pulse encompassing
exactly $n_{\rm c}$ optical cycles.
As in Figs.\
\ref{Fig:newfig1} and \ref{Fig:newfig2}
the carrier wavelength is 800 nm; however here $F_0=0.3$ a.u., corresponding
to a peak intensity of about $3.2 \times 10^{15}$ W cm$^{-2}$, and
$\varphi=0.23$.
Solid black curve: $n_{\rm c}=4$. Solid red curve: $n_{\rm c}=5$.
Dotted blue curve: $n_{\rm c}=6$. Solid green curve: $n_{\rm c}=10$.
Dashed black curve: $n_{\rm c}=15$.
}
\label{Fig:newfig3}
\end{figure*}
Increasing the pulse duration increases
the number of saddle times contributing significantly to the ionization
probability. The impact of this change on the structure of the angle-integrated
energy spectrum is illustrated by 
Fig.\ \ref{Fig:newfig3}.
The results shown in this figure were calculated for the same system
as in Fig.\ \ref{Fig:newfig2} but for a smaller intensity 
and for a single value of
the carrier-envelope phase ($\varphi=0.23$ in Fig.\ \ref{Fig:newfig3}).
At the intensity considered, ionization occurs almost entirely in the vicinity 
of the maximum of the pulse envelope. As only one saddle time is important,
that closest to $t=0$, the spectrum is almost structureless.
The other saddle times become more significant for longer pulse durations.
As a result, the peaks are more
contrasted for 5-cycle pulses, 
very obvious for 6- and, particularly,
10-cycle pulses, and then
tend to split into subpeaks for still longer pulses.

\begin{figure*}
\includegraphics{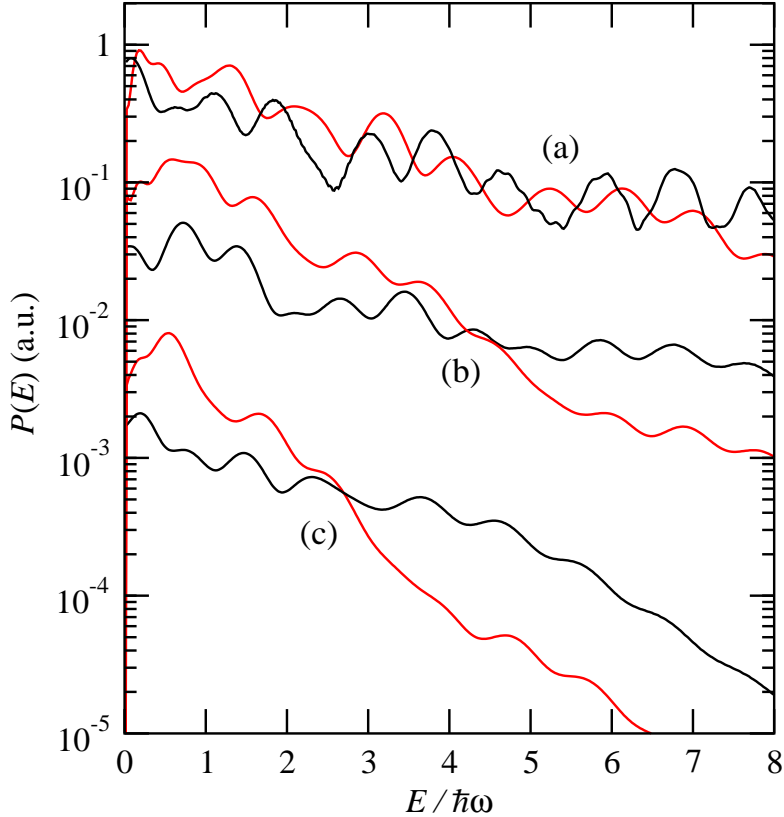}
\caption{(Color online) The angle-integrated probability of ionization
from the ground
state of atomic hydrogen by a cos$^2$ pulse encompassing
exactly 4 optical cycles.
The carrier wavelength is 800 nm and $\varphi=0$. The peak intensity
of the pulse is (a) $2\times 10^{14}$ W cm$^{-2}$, 
(b) $1\times 10^{14}$ W cm$^{-2}$ or
(c) $5\times 10^{13}$ W cm$^{-2}$.
Black curves: Spectrum obtained by solving the time-dependent
Schr{\"o}dinger equation {\it ab initio}.
Red curves: Predictions of the strong field approximation.
}
\label{Fig:newfig7}
\end{figure*}
Interestingly, the structures discussed above 
seem to subsist down to much lower intensities. Both {\it ab initio} and
SFA calculations in atomic hydrogen
for 4-cycle 800 nm pulses
yield angle-integrated energy spectra modulated by near-periodic
maxima separated by about $\hbar\omega$, down to intensities as low 
as $5\times 10^{13}$ W cm$^{-2}$ (Fig.\ \ref{Fig:newfig7}). Although
here the SFA results
are not as close in agreement with the {\it ab initio} results
as in the case of Fig.\ \ref{Fig:newfig6}, they are still
similar in many of their details. 
The oscillations marking the
{\it ab initio} spectra can thus be interpreted as arising primarily from
the interference between saddle times.
There is no indication of resonance structures in Fig.\
\ref{Fig:newfig6}, albeit at these
intensities Stark-shift induced resonances are prominent in picosecond
pulses \cite{Freeman}.

\section{Conclusions}
To conclude, we have shown that for direct (non-recollisional)
ionization or detachment
of atoms or ions by intense few-cycle pulses, the low-energy
part of the angle-integrated
energy spectrum can be modulated by an almost periodic succession of peaks
and troughs. This modulation can be traced to
an energy-dependent interference between the
saddle times of the modified classical action. Depending on the duration and
peak intensity of the pulse, these peaks and troughs may appear either for 
carrier-envelope phases close to zero only or for a wider range of phases.
They are found in {\it ab initio} time-dependent calculations as well as in 
calculations based on the strong field approximation. While much of the
calculations presented in this paper are for the case of an helium atom
or an He$^+$ ion exposed to a super-intense pulse, a similar modulation 
is also observed in atomic hydrogen
at intensities as low as $5 \times 10^{13}$ W cm$^{-2}$.

\begin{acknowledgments}
This work was supported by the A*STAR Computational Resource Centre
through the use of its high performance computing facilities.
Computers financed by EPSRC have also been used to obtain some 
of the numerical results presented here.
The authors thank B.\ Piraux for having provided
programs the calculation of the {\it ab initio} results presented here
was based on.

\end{acknowledgments}

\appendix*
\section{}

Here we comment on the
key numerical issue in the calculation of the
ionization probability
within the approach adopted in this work, which is
the evaluation of the integral of $\exp[iS({\bf p},t)]$
over time.
As is mentioned in Section \ref{sect:theory}, the usual way of dealing
with this integral is to reduce
Eq.\ (\ref{eqn:Amod3}) to Eq.\ (\ref{eqn:As}).
However, a direct numerical intergration,
not relying on this approximation,
can also be contemplated. Special quadradure methods have then to be used,
at least for intense pulses, due to the highly oscillatory nature 
of the integrand. Classical methods such as Gauss quadratures or the 
Simpson method converge, but at the cost of a large number of sampling points,
which makes them time consuming and inefficient.

We have experimented with a direct integration method based on an approach
proposed by Levin \cite{Levin} and further developed by 
Evans and Webster \cite{Evans}. Levin's idea is to make the ansatz
\begin{equation}
\int f(x)\exp[iq(x)]\,dx = y(x) \exp[iq(x)]
\end{equation}
and, given the functions $f(x)$ and $q(x)$, obtain a differential
equation for the unknown function $y(x)$. Levin showed that the relevant
solution of this equation
can be calculated by a collocation method using a polynomial
basis (the choice of the basis eliminates the 
undesired solutions, which are more oscillatory than 
the desired solution). However, this method becomes numerically
unstable if the number of basis functions is excessively
increased in an effort to improve precision.
As argued by Evans and Webster \cite{Evans},
using
Chebyshev polynomials to form the collocation
basis alleviates this
problem of numerical stability. Even with this improvement, however,
 the approach still suffers
from another limitation, which is that the systems of coupled
linear equations which need to be solved are excessively large for
long integration intervals.

We found that this latter limitation can be turned round by subdividing the
integration interval into smaller subintervals such that
a relatively small
Chebyshev basis of 10 to 20 polynomials is sufficient within each subinterval.
The subdivision can be automated into an adaptative algorithm
which subdivides the intervals until a convergence criteria is met. We noticed
that this method is considerably faster than a trapezoidal quadrature of
a same degree of accuracy
for the intense pulses considered in much of this work.

\begin{figure*}
\includegraphics{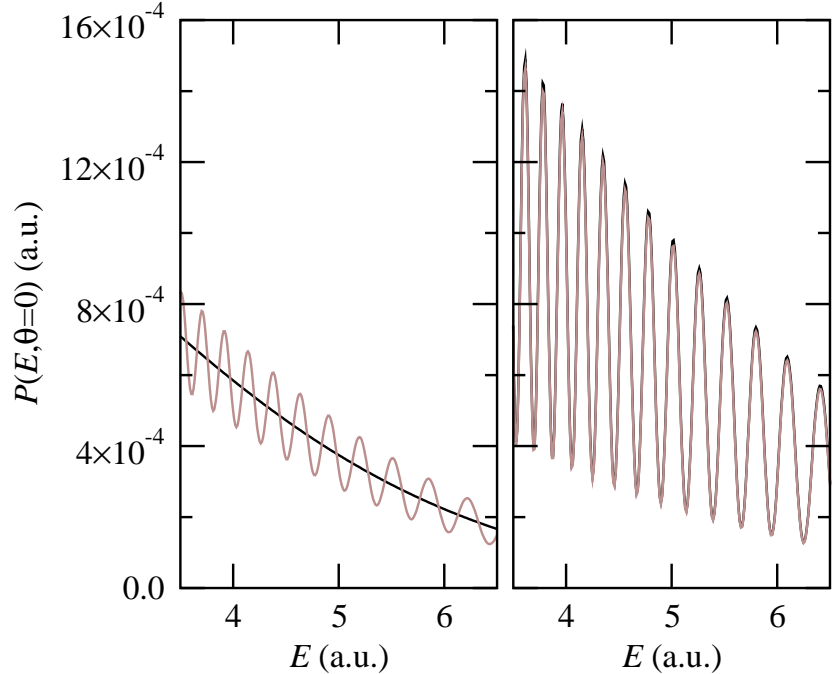}
\caption{(Color online) $P(E,\theta)$ at $\theta=0$ (a) for a 2-cycle
pulse, (b) for a 4-cycle pulse. In both (a) and (b) the pulse
peak intensity is $5\times 10^{15}$ W cm$^{-2}$, the carrier wavelength
is 800 nm, $\varphi=0$, the amplitude envelope is a $\cos^2$ function,
and the target is He$^+$. Light brown curves: results
obtained by integrating 
$\exp[iS({\bf p},t)]$ along the real axis with the end point contributions
removed to first order. Black curves: prediction of Eq.\ (\ref{eqn:As}).
}
\label{Fig:newfig8}
\end{figure*}
However, care should be taken that the end point contributions of the 
integration interval do not affect the resulting ionization probability
(as noted in Section \ref{sect:theory},
these contributions are physically irrelevant).
As an example, the ionization probability predicted by Eq.\ (\ref{eqn:As})
is compared in Fig.\ \ref{Fig:newfig8}
 to that predicted by Eq.\ (\ref{eqn:Amod3})
with the integral performed as explained in the previous 
paragraph. The former decreases monotonically in the case of a 2-cycle
pulse, for which only one saddle time (that closest to $t=0$) contributes 
significantly, while it oscillates in the case a 4-cycle pulse due to the 
interference between several saddle times. The small difference between
these results and the prediction of Eq.\ (\ref{eqn:Amod3})
noticeable in panel (b)
of the figure is indicative of the accuracy of the saddle point method in
this case. However, 
for the still shorter pulse considered in panel (a),
the higher-order
contributions of the end points $t_{\rm i}$ and $t_{\rm f}$
to the integral of $\exp[iS({\bf p},t)]$ are not negligible, because
in our model the
field varies more abruptly at these end points,
and these contributions produce spurious
oscillations in the ionization probability.

This last difficulty can be avoided by making $t$ complex
and deforming the integration
contour into a line running from $t_{\rm i}$ to $t_{\rm i}+i{\cal C}$ where
${\cal C}$ is a positive constant defined below,
from $t_{\rm i}+i{\cal C}$ to $t_{\rm f}+i{\cal C}$, and finally
from $t_{\rm f}+i{\cal C}$ to $t_{\rm f}$. Taking ${\cal C}$ equal to
the imaginary part of the saddle time closest to the real axis makes the
non-end-point contributions of the integrals from $t_{\rm i}$ to $t_{\rm i}+i{\cal C}$ and
from $t_{\rm f}+i{\cal C}$ to $t_{\rm f}$ negligibly small compared to the
integral from $t_{\rm i}+i{\cal C}$ to $t_{\rm f}+i{\cal C}$. 
Their end-point contributions can be relatively large,
but they have no physical
meaning and they can be completely removed by integrating only over the
line running from $t_{\rm i}+i{\cal C}$ to $t_{\rm f}+i{\cal C}$. Along this
line $\exp[iS({\bf p},t)]$ varies slowly, instead of oscillating rapidly
as in the original integral, which makes the numerical quadrature
unproblematic. 

Finally, we comment on the saddle point approximation, Eq.\ (\ref{eqn:As}).
It could be expected that with increasing intensity, and therefore with
increasing values of $S({\bf p},t)$, saddle integration would become more
accurate. However, this is not the case. The reason for this is revealed
by examining the cubic term in the Taylor expansion
of $S({\bf p},t)$ about a saddle time. Making the same approximation as that 
leading to Eq.\ (\ref{eq:interf}) gives,
at a saddle time $t_j$,
\begin{eqnarray}
S''({\bf p},t=t_j)
 &\approx& i |E(t_{0j})|\sqrt{2I_p+p_\perp^2}, \label{eqn:secondorder} \\
S'''({\bf p},t=t_j)
 &\approx& |E(t_{0j})|^2. \label{eqn:thirdorder}
\end{eqnarray}
Clearly, the term in $S'''({\bf p},t_j)$ in the Taylor expansion increases
with intensity faster than that 
in $S''({\bf p},t_j)$, and may therefore
become important for strong enough pulses.
However, the ordinary saddle time method
takes only the latter into account.

Including the cubic
term has been considered previously \cite{Ortner}; however, we are
not aware that the resulting expression of the ionization amplitude in
terms of Airy functions have been used in calculations of the probability
of ionization in few-cycle pulses.
The saddle point method is easily generalized to include the cubic dependence,
though, by making use of the formula
\begin{equation}
\begin{split}
\int_{-\infty}^{\infty} \exp  \left(-{a \over 2}x^2  + i {b\over 6}x^3\right)
& dx = \\ 
 {\pi}\left({16 \over b}\right)^{1/3} & \exp\left(a^3 \over 3b^2\right)
\mbox{Ai}\left[ \left({a^6 \over 4b^4}\right)^{1/3}\right].
\end{split}
\label{eqn:Airy}
\end{equation}
Here
$a \equiv S''({\bf p},t_j)/i$ and $b \equiv S'''({\bf p},t_j)$.
Using the asymptotic form of the Airy function Ai \cite{Abramowitz},
this relation reduces in the limit $b\rightarrow 0$ to
the familiar equation
\begin{equation}
\int_{-\infty}^{\infty} \exp \left(-{a \over 2}x^2\right)
dx = 
 \sqrt{2\pi \over a},
\end{equation}
which underpins the ordinary saddle point method.
We note with Ortner and Rylyuk \cite{Ortner} that
the left-hand side of Eq.\ (\ref{eqn:Airy}) is formally divergent
in applications to the SFA, unless 
the approximation (\ref{eqn:thirdorder}) is made, as $S'''({\bf p},t)$ 
is normally complex at complex saddle times.
Nonetheless, its right-hand side is well defined even for complex values
of $b$.
The usual saddle point result of Eq.\ (\ref{eqn:As}) can thus be improved by 
replacing the factor of $[2\pi i / S''({\bf p},t_j)]^{1/2}$ by the 
right-hand side of Eq.\ (\ref{eqn:Airy}). In practice, the additional
cost is small since fast library routines are available to compute
the Airy function of a complex argument.

\begin{figure*}
\includegraphics{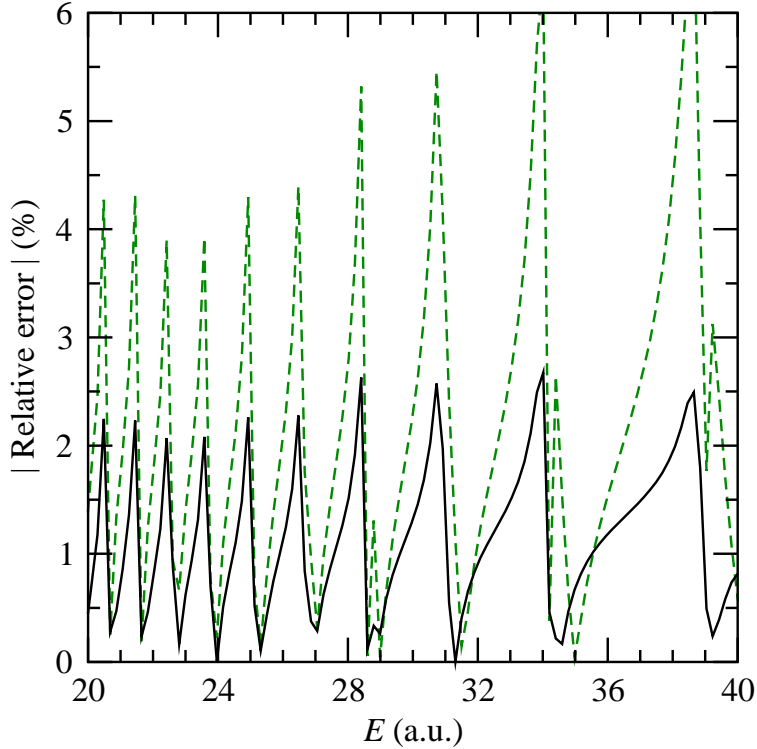}
\caption{(Color online) The relative difference between the value of
$P(E,\theta)$ calculated by direct integration of $\exp[iS({\bf p},t)]$
and that calculated by saddle point integration. 
Dashed curve: the usual saddle point method.
Solid curve: the improved saddle point method described in the text.
These results refer to the case of an He$^+$ ion exposed to an 800 nm, 
4-cycle cos$^2$ pulse of 8.8$\times 10^{15}$ W cm$^{-2}$ peak intensity
with $\varphi=0$.
Here $\theta=0$ and the range of energy
considered is close to the classical cutoff of $2 U_p$ ($2 U_p = 38.5$ a.u.).
}
\label{Fig:newfig9}
\end{figure*}
The improvement in accuracy offered by this method should not be expected
to be significant
for low peak intensities but can be noticeable for ultra intense
pulses.
For instance, for the case of a 800 nm pulse of almost $9 \times
10^{15}$ W cm$^{-2}$ peak intensity, using
Eq.\ (\ref{eqn:Airy}) systematically halves the error on the saddle point 
integration in the higher energy part of the direct ionization spectrum
(Fig.\ \ref{Fig:newfig9}). The error is also generally reduced in the lower
end of the spectrum, although the trend is not as systematic.


\begin{thebibliography}{99}

\bibitem{Agostini}
P.\ Agostini, F.\ Fabre, G.\ Mainfray, G.\ Petite, and N.\ K.\ Rahman,
Phys.\ Rev.\ Lett.\ {\bf 42}, 1127 (1979).
%
\bibitem{Beckerrev}
W.\ Becker, F.\ Grasbon, R.\ Kopold, D.\ B.\ Milosevic, G.\ G.\ Paulus,
and H.\ Walther, Adv.\ At.\ Mol.\ Opt.\ Phys.\ {\bf 48}, 35 (2002).
%
\bibitem{Petite}
E.g., G.\ Petite, P.\ Agostini,
 and H.\ G.\ Muller, J.\ Phys.\ B {\bf 21}, 4097 (1988).
%
\bibitem{Freeman}
R.\ R.\ Freeman, P.\ H.\ Bucksbaum, H.\ Milchberg, S.\ Darack,
D.\ Schumacher, and M.\ E.\ Geusic, Phys.\ Rev.\ Lett.\ {\bf 59}, 1092
(1987).
%
\bibitem{Paulus}
G.\ G.\ Paulus, W.\ Nicklich, H.\ Xu, P.\ Lambropoulos, and H.\ Walther,
Phys.\ Rev.\ Lett.\ {\bf 72}, 2851 (1994).
%
\bibitem{lowen}
C.\ I.\ Blaga, F.\ Catoire, P.\ Colosimo, G.\ G.\ Paulus, H.\ G.\ Muller,
P.\ Agostini, and L.\ F.\ DiMauro, Nature Phys.\ {\bf 5}, 335 (2009);
F.\ Catoire, C.\ I.\ Blaga, E.\ Sistrunk, H.\ G.\ Muller, P.\ Agostini,
and L.\ F.\ DiMauro, Laser Phys.\ {\bf 19}, 1574 (2009);
W.\ Quan, Z.\ Lin, M.\ Wu, H.\ Kang, H.\ Liu, X.\ Liu, J.\ Chen,
J.\ Liu, X.\ T.\ He, S.\ G.\ Chen, H.\ Xiong, L.\ Guo, H.\ Xu,
Y.\ Fu, Y.\ Cheng, and Z.\ Z.\ Xu, Phys.\ Rev.\ Lett.\ {\bf 103},
093001 (2009);
C.\ Y.\ Wu, Y.\ D.\ Yang, Y.\ Q.\ Liu, Q.\ H.\ Gong, M.\ Wu, X.\ Liu,
X.\ L.\ Hao, W.\ D.\ Li, X.\ T.\ He, and J.\ Chen, Phys.\ Rev.\ Lett.\
{\bf 109}, 043001 (2012);
H.\ Liu, Y.\ Liu, L.\ Fu, G.\ Xin, D.\ Ye, J.\ Liu, X.\ T.\ He, 
Y.\ Yang, X.\ Liu, Y.\ Deng, C.\ Wu, and Q.\ Gong,
Phys.\ Rev.\ Lett.\ {\bf 109}, 093001 (2012).
%
\bibitem{Book}
See, e.g., C.\ J.\ Joachain, N.\ J.\ Kylstra and R.\ M.\ Potvliege,
{\it Atoms in Intense Laser Fields} (Cambridge University Press,
Cambridge, 2012).
%
\bibitem{Keldysh}
L.\ V.\ Keldysh, Zh.\ \'{E}ksp..\ Teor.\ Fiz.\ {\bf 47}, 1945 (1964)
[Sov.\ Phys.\ JETP {\bf 20}, 1307 (1965)]. 
%
\bibitem{Kuchiev}
G.\ F.\ Gribakin and M.\ Yu.\ Kuchiev, Phys.\ Rev.\ A {\bf 55},
3760 (1997).
%
\bibitem{CCRMP}
C.\ C.\ Chiril\u{a} and R.\ M.\ Potvliege,
Phys.\ Rev.\ A {\bf 71}, 021402 (2005).
%
\bibitem{thesis}
A preliminary account of part of the results presented here has been
given in
C.\ C.\ Chiril\u{a}, {\it Analysis of the Strong Field Approximation
for Harmonic Generation and Multiphoton Ionisation in Intense Ultrashort
Laser Pulses}, PhD Thesis, Durham University (2004).
%
\bibitem{cep}
If we approximate the corresponding electric field by
$-F_0 \chi(t) \hat{\bfg{\epsilon}} 
\cos (\omega t + \varphi)$, thus neglect the variation
of the envelope function $\chi(t)$ when differentiating ${\bf A}(t)$,
then $\varphi$ is the phase-angle difference between
a peak of the $\cos(\omega t + \varphi)$ carrier wave
and the maximum of the electric field envelope.
%
\bibitem{Dondera}
See, e.g., M.\ Dondera, H.\ G.\ Muller, and M.\ Gavrila,
Laser Phys.\ {\bf 12}, 415 (2002).
%
\bibitem{Krainov}
V.\ P.\ Krainov and B.\ Shokri,
Zh.\ \'{E}ksp.\ Teor.\ Fiz.\ {\bf 107}, 1180
(1995) [JETP {\bf 80}, 657 (1995)];
V.\ P.\ Krainov,
J.\ Opt.\ Soc.\ Am.\ B {\bf 14}, 425 (1997).
See also
 A.\ M.\ Perelomov and V.\ S.\ Popov,
 Zh.\ \'{E}ksp.\ Teor.\ Fiz.\ {\bf 52}, 514 (1967)
 [Sov.\ Phys.\ JETP {\bf 25}, 336 (1967)].
%
\bibitem{boundary2}
The end-point contributions also vanish in the case
of a monochromatic field if $t_{\rm f}-t_{\rm i}$ is
an integer multiple of the field period and the integral
is evaluated for a value of $p^2/2$ satisfying conservation
of energy, as in that case 
$\exp[iS({\bf p},t)]$ and all its derivatives have the same values
at $t_{\rm i}$ as at $t_{\rm f}$.
%
\bibitem{PPT}
 A.\ M.\ Perelomov, V.\ S.\ Popov, and M.\ V.\ Terent'ev,
 Zh.\ \'{E}ksp.\ Teor.\ Fiz.\ {\bf 51}, 309 (1966)
 [Sov.\ Phys.\ JETP {\bf 24}, 207 (1967)].
%
\bibitem{popovreview}
V.\ S.\ Popov, Usp.\ Fi.\ Nauk {\bf 174}, 921 (2004)
[Phys.\ Usp.\ {\bf 47}, 855 (2004)].
%
\bibitem{velgauge}
As long as the end-point contributions can be neglected and that 
no allowance is made for the Coulomb interaction at the tunneling stage
of the process,
the ionization amplitude is also proportional to the integral
of $\exp[iS({\bf p},t)]$ in the Faisal-Reiss (velocity gauge) formulation
of the SFA, although with a different overall factor 
\cite{Book}.
%
\bibitem{Bleistein}
N.\ Bleistein and R.\ A.\ Handelsman,
{\it Asymptotic Expansions of Integrals} (Dover, New York, 1986).
%
\bibitem{methodTD}
The {\it ab initio} calculations follow the method described by
E.\ Huens, B.\ Piraux, A.\ Bugacov, and M.\ Gajda,
Phys.\ Rev.\ A {\bf 55}, 2132 (1997).
%
\bibitem{energyconservation}
The argument is of course the same as that
leading to the conservation of energy condition
$E_N = N\omega - (I_p+U_p)$, where $U_p$ is the ponderomotive energy,
for the case of a monochromatic field. Here the comb of peaks is only
approximately periodic: the position of the peaks is less and less well 
predicted by Eq.\ (\ref{eqn:Eq23}) as $E$ increases.
%
\bibitem{explanation}
Because of the dependence in $p_\perp$
of the right-hand side of Eq.\ (\ref{eq:condition}),
the saddle times have a larger imaginary part for $\theta \approx 90$ deg
than for angles closer to 0 or 180 deg. Correspondingly, 
$I_{jk}$ is suppressed around $\theta = 90$ deg for any pair of saddle 
times. The regions $\theta \approx 0$ and $\theta \approx 180$ deg also
contribute less to the angle integrated spectrum, these ones because
of the $\sin \theta$ factor in the angular integral.
%
\bibitem{Levin}
D.\ Levin, Math.\ Comput.\ {\bf 38}, 531 (1982).
%
\bibitem{Evans}
G.\ A.\ Evans and J.\ R.\ Webster, Appl.\ Numer.\ Math.\ {\bf 23}, 205 (1997).
%
\bibitem{Ortner}
J.\ Ortner and V.\ M.\ Rylyuk, J.\ Phys.\ B {\bf 34}, 3251 (2001).
%
\bibitem{Abramowitz}
H.\ A.\ Antosiewicz, in {\it Handbook of Mathematical Functions}, edited by
M.\ Abramowitz and I.\ A.\ Stegun (U.S.\ Government Printing Office, Washington,
1972).

\end{thebibliography}
\end{document}